\documentclass[usenatbib]{mn2e}
\usepackage{amsfonts}
\usepackage{amsmath}
\usepackage{graphicx}
\usepackage{natbib}

\def\apjl{Astrophys.\ J.\ Lett.}
\def\mnras{Mon.\ Not.\ R.\ Astron.\ Soc.}

\def\aap{Astron.\ Astrophys.}
\def\apj{Astrophys.\ J.}

\def\apjs{Astrophys.\ J. Supp.}

\def\prd{Phys.\ Rev.\ D}

\def\physrep{Phys. Rep.}

\title[Redshift space dark matter PDF in the $f_{nl}$ model]
      {The nonlinear redshift space probability distribution 
       function in models with local primordial non-Gaussianity}

\author[T. Y. Lam, V. Desjacques \& R. K. Sheth]
 {Tsz Yan Lam$^{1,2}$\thanks{E-mail:tszyan.lam@ipmu.jp, dvince@physik.uzh.ch, shethrk@physics.upenn.edu},
  Vincent Desjacques$^3$\footnotemark[1] \& Ravi K. Sheth$^1$\footnotemark[1] \\
 $^1$ Center for Particle Cosmology, University of Pennsylvania, 
 209 S. 33rd Street, Philadelphia, PA 19104, USA\\
$^2$ Institute for the Physics and Mathematics of the Universe, University of Tokyo, Kashiwa, Chiba 277-8568, Japan\\
$^3$ Institute for Theoretical Physics, University of Z\"urich, 
 Winterthurerstrasse 190, CH-8057 Z\"urich, Switzerland}

\newcommand{\bm}[1]{{\mbox{\boldmath $#1$}}}

\begin{document}
\pagerange{\pageref{firstpage}--\pageref{lastpage}}

\maketitle

\label{firstpage}

\begin{abstract}
We use the ellipsoidal collapse approximation to investigate the nonlinear
redshift space evolution of the density field with primordial non-Gaussianity 
of the local $f_{nl}$-type.  
We utilize the joint distribution of eigenvalues of the initial non-Gaussian 
shear field and evaluate the evolved redshift space probability distribution 
function (PDF). 
It is shown that, similar to the real space analysis, 
the underdense tail of the nonlinear redshift space PDF 
differs significantly from that for Gaussian initial conditions.
We also derive the lowest order correction of the Kaiser's formula
in the presence of a non-zero $f_{nl}$.
\end{abstract}

\begin{keywords}
methods: analytical - dark matter - large scale structure of the universe 
\end{keywords}

\section{Introduction}
Cosmological probes of primordial non-Gaussianity have recently attracted 
much attention because of their potential ability to discriminate between 
different inflationary models 
\citep[e.g.,][and references therein]{bko08,kp08,st08}.
Constraints on  primordial non-Gaussianity mainly come from 
the CMB \citep{cmb5yr,hikageetal08,yw08,mhlm08,rsphmfnl} and 
large scale structures in the Universe 
\citep{kst99,mvj00,ssz04,sk07,is07,fnlverde,dalaletal08,mv08,cvm08,
at08,shshp08,mcdonald08,tkm08,slosar08,grossi08,kvj08,fnlvincent,pphfnl08,
lamshethfnl,grossinfm,lsdfnl}.

This paper is concerned with one particular measure of large scale 
structures:  the probability that a cell of volume $V$, placed at 
random in the nonlinear redshift space density field, contains 
a certain amount of mass (or, equivalently, is denser than the 
background by a certain amount).  This statistic is known as the 
nonlinear redshift space probability distribution function (PDF).
Our goal is to estimate this distribution for scales as small as a 
few Mpc in the local non-Gaussian model, in which the primordial 
perturbation potential is 
\begin{equation}
 \Phi = \phi + f_{nl}(\phi^2 - \langle \phi^2\rangle).
 \label{eqn:fnl}
\end{equation}
Here, $\Phi$ is the Bardeen potential, 
$\phi$ is a Gaussian potential field and $f_{nl}$ is the
nonlinear quadratic parameter. 
The right-hand side of eq.~(\ref{eqn:fnl})
shows the first two terms of an (infinite) Taylor series in $\phi$.
However, since $|\phi|\sim 10^{-5}$, one usually ignores higher order
corrections, and commonly refers to this simplified model as the
$f_{nl}$ model.
This definition of $\Phi$ is consistent with most of the recent
studies on the local $f_{nl}$ model \citep[but our earlier studies,]
[defined $\Phi$ as the Newtonian potential]{lamshethfnl,lsdfnl}.

Our approach is based on previous work which develops the formalism 
needed for estimating the evolution of the density PDF from Gaussian initial 
conditions in real and redshift space \citep{lamshethreal, lamshethred}.  
The evolved PDF depends on the collapse dynamics and the statistical
properties of the initial density field.  Recently,
\citet{lamshethfnl} used the fact that only the initial conditions are
affected by primordial non-Gaussianity to model the evolution of the
real  space nonlinear PDF for the local non-Gaussian model.  Their
approach provided good quantitative agreement with measurements  in
numerical simlations.  In what follows, we will assess whether this
is also true in redshift space.

Although it is possible to study the evolution of the redshift space
PDF using perturbation theory methods
\citep{b94,hbcj95,hept,ptreview},  this has, somewhat surprisingly not
been extended to the local non-Gaussian  model.  Thus, it is not
obvious how Kaiser's formula relating the variance (the second
moment of the PDF) in real and redshift space \citep{kaiser} 
is modified when the initial conditions are non-Gaussian.  
\citep[Although Kaiser's original derivation makes no explicit assumption 
about Gaussianity, the Gaussian assumption plays an important role 
in other derivations of his formula, e.g.,][]{fisher95, ecpdf}.  
Our approach is quite different from Kaiser's, as it is  based on an 
approximate model for the dynamics -- the  ellipsoidal collapse 
model -- which reduces to perturbation theory  at early times 
\citep{bm96}, but allows one to study more nonlinear structures 
\citep{smt01,d07}.
\cite{lamshethred} showed that the  ability to probe deeper into the
nonlinear regime, using a dynamical  model that does {\em not} assume
spherical symmetry, was crucial for  modelling the PDF, especially in
redshift space.  However,  implementing this approach requires
knowledge of the initial shear  field.  For Gaussian initial
conditions, this has been known for  some time \citep{grf}, but how
Doroshkevich's formulae are modified  for the local $f_{nl}$ model has
been shown only recently \citep{lsdfnl}.  Hence, we now have the
necessary ingredients to study the redshift  space PDF.

Properties of the initial shear field in the local non-Gaussian model
are briefly reviewed in Section~\ref{section:plambdas}.  The dynamics
of ellipsoidal collapse and the calculation  of the nonlinear redshift
space PDF are described in  Sections~\ref{section:ellrho}
and~\ref{section:zspacePDF}.  We compare our model predictions with
numerical simulations in  Section~\ref{section:sims}.  We summarize
our results in Section~\ref{section:discussion}.  A perturbative
treatment of our model is given in  Appendix~\ref{section:PT}; this
shows explicitly that Kaiser's  formula holds to lowest order in
$f_{nl}$, but that at higher  order, it is modified.


\section{The redshift space density PDF in the local non-Gaussian model}
\label{section:def}
Let us define the nonlinear overdensity of a region of volume $V$ 
containing mass $M$ by 
\begin{equation}
\rho \equiv 1 + \delta = \frac{M}{\bar{\rho}V},
\end{equation}
where $\bar{\rho}$ is the mean density.  We will use $\rho_s$ to 
denote the corresponding quantity in redshift (rather than real) space.  
This section studies the expected dependence of the PDF of $\rho_s$ on 
the value of $f_{nl}$, when the primordial potential is given by 
equation~(\ref{eqn:fnl}).

To proceed, we use the assumptions made when dealing with Gaussian initial 
distributions ($f_{nl}=0$): 
there is a local mapping from the eigenvalues $\lambda_j$ of the 
initial deformation tensor to the nonlinear overdensity $\rho_s$ 
(see section~\ref{section:ellrho}); 
and statistics on the smoothing scale $V$ at the present time are 
related to statistics on a different smoothing scale in the initial 
conditions --  the relevant initial smoothing scale is the one which 
contains the same mass (so it is larger for overdense cells, and 
smaller for underdense cells).  

Therefore, the nonlinear redshift space PDF of $\rho_s$ is given by
\begin{equation}
 \rho_s\, p(\rho_s|V){\rm d}\rho_s = \int {\rm d}{\bm \lambda}\,
     {\rm d}{\bm e}\,p({\bm \lambda}|\sigma)\,
      \delta_{\rm D}\left[\rho_s = \rho_s'({\bm \lambda},{\bm e})\right],
 \label{eqn:ecpdf}
\end{equation}
where $\rho_s \equiv M/\bar{M}$ ($\bar M\equiv\bar\rho V$ is the 
average mass in cells of size $V$), 
${\bm\lambda}$ denotes the 3 eigenvalues 
(our convention is to have $\lambda_1 \geq \lambda_2 \geq \lambda_3$)
of the initial $3\times 3$ deformation tensor when smoothed on 
scale $M$ (not $V$), 
$\sigma^2$ denotes the variance of the initial density fluctuation 
field on this smoothing scale 
(the initial density fluctuation $\delta_l$ is defined by 
$\delta_l\equiv {\rm Tr}\,{\bm \lambda}$), 
$\rho_s'({\bm \lambda},{\bm e})$ is the local mapping from the 
initial field to the evolved density given by the ellipsodial 
collapse model (spherical evolution models assume that the mapping is 
driven by the initial density $\delta_l$ only), 
and ${\bm e}$ represents the rotation vector from the 
line-of-sight direction to the principle axis of the ellipsoid.
Equation~(\ref{eqn:ecpdf}) has the same form as equation~(8) of 
\citet{lamshethred} but, in our case, $p({\bm \lambda}|\sigma)$ is 
the joint distribution of the initial eigenvalues $\lambda_j$ in 
the $f_{nl}$ model rather than in the Gaussian model.  

Before we compute $p({\bm \lambda}|\sigma)$, note that 
equation~(\ref{eqn:ecpdf}) does not guarantee a properly normalized 
PDF.  To ensure the correct normalization, we set $\rho' = N \rho$ and 
$\rho'^2 p(\rho') = \rho^2 p(\rho)$ where $N$ is chosen so that 
both $\int {\rm d}\rho'\, p(\rho')$ and $\int {\rm d}\rho'\,\rho' p(\rho')$ 
equal unity \citep{lamshethreal}.

\subsection{Initial conditions in the $f_{nl}$ model}
\label{section:plambdas}
Let $p({\bm \lambda}|\delta_l,\sigma)$ denote the distribution of the 
$\lambda_j$ at fixed $\delta_l$, and let $p_0({\bm \lambda}|\sigma)$ 
and $p_0({\bm \lambda}|\delta_l,\sigma)$ denote the corresponding quantities 
when $f_{nl}=0$, i.e., for Gaussian initial conditions.  (Note that 
this means $p_0(\delta_l)$ is a Gaussian.)  One of the main results 
of \cite{lsdfnl} was to show that 
\begin{eqnarray}
 p({\bm \lambda}|\delta_l,\sigma) &=& p_0({\bm \lambda}|\delta_l,\sigma)
 \nonumber\\
  &=& \frac{3^4/4}{\Gamma(5/2)} 
               \left(\frac{5}{2\sigma^2}\right)^{5/2}%
              \exp\left(-\frac{5\delta_l^2}{2\sigma^2} +  %
                       \frac{15I}{2\sigma^2}\right)  \nonumber \\
  && \qquad \times\ 
     (\lambda_1-\lambda_2)(\lambda_2-\lambda_3)(\lambda_1-\lambda_3),
\end{eqnarray} 
where the final expression for $p_0$ is from \cite{grf}, 
and $I$ is the sum of the three permutations of $\lambda_i\lambda_j$ 
where $i\ne j$. 
Therefore the joint distribution of $\lambda_j$ in the $f_{nl}$ model is
\begin{equation}
 p({\bm \lambda}|\sigma)
   =  p(\delta_l|\sigma)\,p({\bm \lambda}|\delta_l,\sigma) %
   = p(\delta_l|\sigma)\,p_0({\bm \lambda}|\delta_l,\sigma),
\label{eqn:pdflambdas}
\end{equation}
where $p(\delta_l|\sigma)$ is the distribution of the linear overdensity 
in the $f_{nl}$ model, and $p_0({\bm \lambda}|\delta_l,\sigma)$ is 
really a function of $\lambda_i/\sigma$ and $\delta_l/\sigma$.
For the $f_{nl}$ values of current interest, $p(\delta_l|\sigma)$ is 
only weakly non-Gaussian, so it can be approximated by the Edgeworth 
expansion \citep[e.g.,][who also discuss the limitations of this 
approximation]{lamshethfnl}.  Hence, the joint distribution of 
$\lambda_j$ is
\begin{equation}
 p({\bm \lambda}|\sigma) = 
  \left[1 + \frac{\sigma S_3}{6}H_3(\delta_l/\sigma)\right]\,
   p_0({\bm \lambda}|\sigma),
\label{eqn:plambda}
\end{equation}
where $H_3(\delta_l/\sigma) = (\delta_l/\sigma)^3 - 3(\delta_l/\sigma)$ is 
the Hermite polynomial.  The dependence on $f_{nl}$ is encoded in 
the skewness parameter $\sigma S_3$ \citep[e.g.,][and note that our 
convention means that $S_3$ is of same sign to $f_{nl}$]{ssz04}.   
As a result, equation~(\ref{eqn:ecpdf}) becomes 
\begin{align}
\rho_s p(\rho_s|V)\,{\rm d}\rho_s = & 
   \int {\rm d}{\bm \lambda}\, {\rm d}{\bm e}%
        \left[1 + \frac{\sigma S_3}{6}H_3(\delta_l/\sigma)\right] \nonumber \\
  &  \quad \times   p_0({\bm \lambda}|\sigma) \,
     \delta_{\rm D} \left[\rho_s = \rho_s'({\bm \lambda},{\bm e})\right].
\label{eqn:ecpdfw}
\end{align}
Except for the term in square brackets (the Edgeworth correction factor), 
the quantity in the integral is the same as in the Gaussian case.  If we 
think of this extra factor as a weight, then the resulting nonlinear 
redshift PDF in the $f_{nl}$ model is just a suitably weighted version 
of that in the Gaussian case.  The weight depends on $\sigma S_3$, i.e. 
on $f_{nl}$. 

We can gain some intuitive understanding of the effect of  a nonzero
$f_{nl}$ as follows.   For $f_{nl} = -100$, $\sigma S_3 < 0$ so that
overdense regions are  suppressed compared to the Gaussian case (the
weight factor is  less than unity), whereas underdense regions are
enhanced  compared to the Gaussian case (the weight factor is larger
than unity).
Finally, note that $|\sigma S_3| \ll 1$; on Mpc scales  $\sigma
S_3\approx -0.03$ for $f_{nl}=-100$, and it is a weakly decreasing
function of scale \citep[e.g.,][]{ssz04}.  This will be important in
what  follows.

\begin{figure*}
\centering
\includegraphics[width =0.7\hsize]{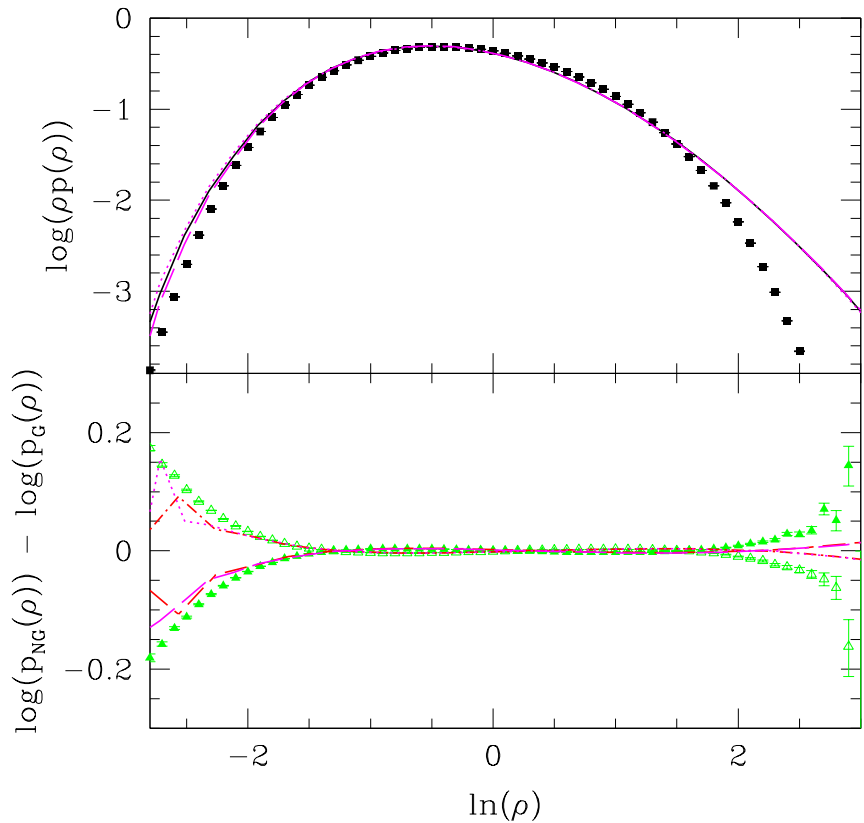}
\caption{Comparisons of the measured PDF with our model for counts in 
         $8h^{-1}{\rm Mpc}$ sphere. 
         The upper panel shows $\log(\rho p(\rho))$ against $\ln(\rho)$ 
         for the measured PDF (solid symbols, $f_{nl}=0$) and the 
         theoretical prediction obtained by evaluating the nonlinear PDF 
	 eq.~(\ref{eqn:ecpdf}) for $f_{nl}= 0$ (black, solid), 
         $-100$ (magenta, dotted), and $100$ (magenta, dashed) respectively.
         The lower panel shows the logarithm of the ratio between the 
         $f_{nl} \neq 0$ and Gaussian counts.
         The filled and empty symbols indicate the measurement for $f_{nl} = 100$ 
	 and $-100$, respectively.
         The predictions obtained by applying the Edgeworth expansion weighting 
         are represented by the dot-dashed (red, $f_{nl} = -100$) and the 
         short-long-dashed (red, $f_{nl} = 100$), respectively.
         }
\label{fig:ecptr8}
\end{figure*}

\subsection{Ellipsoidal collapse and the nonlinear overdensity}\label{section:ellrho}
The next step is to estimate how $\rho_s$ depends on $({\bm \lambda},e)$.  
In real space, the ellipsoidal evolution model sets 
\begin{equation}
\rho_r \equiv \prod_{j=1}^3 \frac{R_j^i}{R_{\rm E}}
       \approx \frac{(1 - \delta_l/3)^3}{(1-\delta_l/\delta_c)^{\delta_c}}
               \prod_{j=1}^3 (1-\lambda_j)^{-1}, \label{eqn:rhor}
\end{equation}
\citep{lamshethreal}, where $R_j^i$ are the initial lengths of 
the patch which is now an sphere of radius $R_{\rm E}$, and 
$\delta_c$ is the critical value of spherical collapse model 
(its exact value depends weakly on cosmology: 
 $\delta_c \approx 1.66$ for the $\Lambda$CDM cosmology for which we 
 show simulation data in the next section).  
In redshift space, the model sets 
\begin{equation}
 \frac{\rho_s}{\rho_r} \approx  
 \left[1-\sum_{k=1}^3\frac{f\left\{R_k^i\lambda_k-A^i_h\delta_l\left[1-(1-\delta_l/\delta_c)^{\delta_c/3-1}\right]/3\right\}}{R_k^i(1-\lambda_k)-A^i_h\left[1-\delta_l/3-(1-\delta_l/\delta_c)^{\delta_c/3}\right]}e_k^2 \right]^{-1},
\label{eqn:ellred}
\end{equation}
\citep{lamshethred}, where
 $f = {\rm d} \ln D/{\rm d} \ln a$ with $D(t)$ the linear growth factor, 
$(e_1,e_2,e_3) = (\cos\psi\sin\theta,\sin\psi\sin\theta,\cos\theta)$, 
and
 $A_h^i\equiv 3/\sum_j (R_k^i)^{-1}$, where the 
 $R_j^i$ are the initial axis lengths.

\begin{figure*}
 \centering
 \includegraphics[width =0.7\hsize]{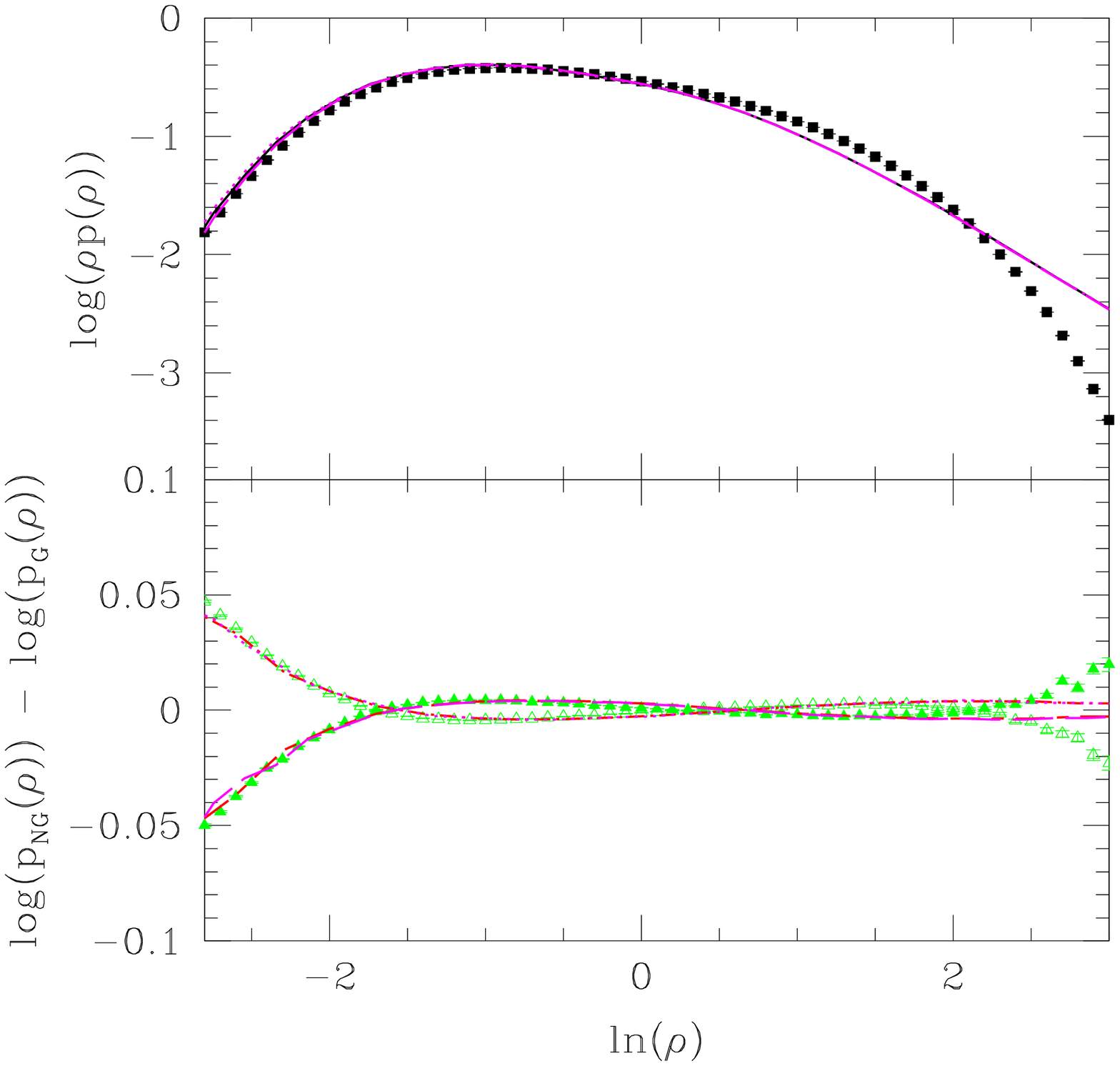}
 \caption{Same as figure~\ref{fig:ecptr8}, but for counts in spheres
          of radius $4h^{-1}{\rm Mpc}$.}
 \label{fig:ecptr4}
\end{figure*}

\subsection{Nonlinear PDF in redshift space}\label{section:zspacePDF}
Equations~(\ref{eqn:ecpdf}) and (\ref{eqn:ellred}) are the bases for 
the computation of the nonlinear redshift space PDF. 
The analysis simplifies considerably if we approximate $\sigma S_3$ 
as a constant for a given Eulerian smoothing scale (recall that the 
scale dependence is rather weak):  
 $\sigma S_3(r_0,\rho_s) \approx \sigma S_3(r_0)$. 
With this assumption we can write equation~(\ref{eqn:ecpdf}) as
\begin{align}
 \rho_s\, p(\rho_s|V){\rm d}\rho_s & = \int {\rm d}{\bm \lambda}\,
     {\rm d}{\bm e}\,p({\bm \lambda}|1)\,
      \delta_{\rm D}\left[\rho_s = \rho_s'(\sigma{\bm \lambda},{\bm e})\right]
     \nonumber\\
  & = \int {\rm d}{\bm \lambda}\, {\rm d}{\bm e}%
        \left[1 + \frac{\sigma S_3}{6}H_3(\delta_l)\right] \nonumber \\
  &  \qquad \times\   p_0({\bm \lambda}|1) \,
     \delta_{\rm D}\left[\rho_s = \rho_s'(\sigma{\bm \lambda},{\bm e})\right].
 \label{eqn:ecpdfrho}
\end{align}
In practice, we construct the PDF by Monte Carlo solution of the
integral.  This is straightforward because the six independent
components of the  deformation tensor $\Phi_{ij} = \partial_i
\partial_j \Phi$ can  be combined in the form
$C=\{x,y,z,\Phi_{12},\Phi_{23},\Phi_{31}\}$,  where
\begin{align}
x = \sum_i \Phi_{ii}, \ 
y = \frac{\Phi_{11} - \Phi_{22}}{2}, \ 
z = \frac{\Phi_{11} + \Phi_{22} - 2\Phi_{33}}{2}.
\end{align}
The reason for doing this is that, to second order in $f_{nl}$, only
$x$ has non-zero  skewness \citep{lsdfnl}.  Therefore, we can draw the
other five parameters from Gaussian  distributions with variance
\begin{equation}
\langle y^2 \rangle = \langle \Phi_{ij}^2 \rangle_{i\neq j}
                    = \frac{\sigma^2}{15} , \quad {\rm and}\quad
\langle z^2 \rangle = \frac{\sigma^2}{5}.
\end{equation}
The parameter $x$ is drawn from an Edgeworth distribution.  
The associated $(\lambda_1,\lambda_2,\lambda_3)$ can be computed 
by solving the eigenvalue problem and the nonlinear redshift PDF is 
then evaluated using equation~(\ref{eqn:ecpdfrho}).

In the next section, we compare this full solution with measurements
in simulations.  Note however that, in the limit $\rho_s-1\ll 1$, the
nonlinear redshift space PDF can be solved perturbatively.
Appendix~\ref{section:PT}  provides details and shows that, to lowest
order, the variance in  the redshift space counts is related to that
in real space by Kaiser's  formula for $f_{nl}=0$; the dependence on
$f_{nl}$ enters at higher  order.

\section{Comparison with simulations}\label{section:sims}
We now compare the predictions of our model with measurements of the 
nonlinear PDF in numerical simulations from \citet{fnlvincent}.  
The numerical simulations followed the evolution of $1024^3$ particles 
in a periodic cube of sides $1600 h^{-1}{\rm Mpc}$.  The background 
cosmology was $\Lambda$CDM with
 $(\Omega_m,\Omega_b,n_s,h,\sigma_8) = (0.279,0.0462,0.96,0.7,0.81)$.

Figure~\ref{fig:ecptr8} compares our model with the measured redshift
space PDF of counts in $8h^{-1}{\rm Mpc}$ spheres.  In the upper
panel, the solid symbols show the PDF for $f_{nl}=0$.  We have not
shown results for $f_{nl}= \pm 100$ in that panel since  they only
slightly differ from the Gaussian case. Instead, the symbols  in the
bottom panel show the fractional deviation in these models  relative
to the Gaussian case.  Filled and open symbols are for $f_{nl} = 100$
and $f_{nl}=-100$,  respectively.   As we can see, a positive $f_{nl}$
slightly skews the  PDF towards overdense regions. Conversely, the
fraction of underdense regions is enhanced for negative $f_{nl}$.  In
this respect, the redshift space PDF shows the same  qualitative
dependence on $f_{nl}$ as the real space PDF,  as expected
(c.f. discussion following equation~\ref{eqn:ecpdfw}).

The dashed, solid, and dotted curves in the top panel show the
predictions for $f_{nl} = 100$, $0$, and $-100$ respectively.   The
differences are small, so the curves appear almost identical,  but the
bottom panel shows that they are indeed slightly different  from one
another, and that our model provides a good description  of the
ratios, except in the high density tail where it  underpredicts the
dependence on $f_{nl}$. Note however that, in this strongly nonlinear
regime, our model drastically overpredicts the Gaussian counts.

We believe we understand why our model is more successful at
predicting the ratio than the counts themselves.  This is because  at
least some of the discrepancy at $\rho_s\gg 1$ arises from the  fact
that the highly nonlinear virial motions within halos will act  to
erase large density contrasts -- these motions are {\em not} part  of
our model.  Figures~1 and~2 in \cite{lamshethred} show that,  for
$f_{nl}=0$, virial motions reduce the $\rho_s\gg 1$ tail, enhance  the
intermediate $\rho_s\sim 2$ region of the PDF, and have almost no
effect on the $\rho_s<0$ regime.  Removing virialized motions within
halos from the measurements substantially reduces the discrepancy
between theory and the simulations at $\rho_s>0$.  Although virial
motions do not depend on $f_{nl}$, their net effect depends upon the
halo mass function. Since the later does depend on $f_{nl}$, we   may
thus expect a slightly stronger suppression when $f_{nl}>0$ (as the
abundance of massive halos is slightly enhanced).  On the other hand,
the real space  PDF, which is also affected by virial motions, has a
more pronounced  high density tail for $f_{nl}>0$. As a result, the
ratio of non-Gaussian  counts to Gaussian counts depends only weakly
on $f_{nl}$. Therefore,  our model can provide a reasonable
description of the ratio even though it  fails at describing the high
$\rho_s$ tail of the Gaussian density PDF.

Figure~\ref{fig:ecptr4} shows a similar comparison on smaller  scales
(spheres of radius $4h^{-1}$Mpc).  The dependence on $f_{nl}$  is
smaller compared to the previous figure. Our model still  provides a
good description of the ratio relative to the $f_{nl}=0$  counts,
except at the highest densities where it overpredicts  the $f_{nl}=0$
counts and underpredicts the ratio.  Note again,  that we expect much
of this discrepancy to be reduced if we were to  remove virial motions
from the simulations.

\section{Discussion}\label{section:discussion}
We used the ellipsoidal evolution model to study the redshift space 
probability distribution function of the nonlinear dark matter density 
field in the local non-Gaussian $f_{nl}$ model. 

A perturbative analysis of the density PDF eq.(\ref{eqn:ecpdf}) shows 
that, at the lowest order, Kaiser's formula still holds in the 
$f_{nl}$ model (although his original derivation does not assume 
Gaussianity explicitly, other derivations of the formula have done 
so, as discussed in the Introduction).  The effects of $f_{nl}\ne 0$ 
appear in the first order corrections to the variance (and higher 
order moments).
One could, therefore, constrain $f_{nl}$ from large scale structure 
by measuring the variance and the higher order moments 
and comparing with the perturbative quantities in Section~\ref{section:PT}
(with some dynamical models to determine $\nu_i$).  

Our approach remains accurate on smaller scales where perturbative 
treatments are not useful.  Simulations  show that the dependence 
on $f_{nl}$ is qualitatively similar to that for the real space PDF: 
for positive $f_{nl}$ (positive $\sigma S_3$) both PDFs skew slightly 
towards overdense regions.  
In addition both show stronger $f_{nl}$ dependence in the underdense 
regions, suggesting that void abundances should be good probes for 
primordial non-Gaussianity \citep[e.g.,][]{kvj08,lsdfnl}.  Our model 
(equation~\ref{eqn:ecpdfw}) captures these trends 
(Figures~\ref{fig:ecptr8} and~\ref{fig:ecptr4}).  Since it is explicitly 
a redshift space calculation, it would be interesting to see if it 
correctly predicts the $f_{nl}$ dependence of the PDF of the flux in 
the Ly-$\alpha$ forest, that has recently been simulated by \cite{viel09}.
This work also provides the foundation for constraining $f_{nl}$ in
future galaxy surveys (e.g. the change in the redshift
space halo/galaxy power spectrum by combining with the scale dependent
halo/galaxy bias \citep{dalaletal08,slosar08,fnlvincent}). 

\section*{Acknowledgements}
V.D. acknowledges support from the Swiss National Foundation under
contract No. 200021-116696/1.  RKS was supported in part by NSF-AST 0908241.



\appendix
\section{Perturbative treatment of redshift space distortions}
\label{section:PT}
Equation~(\ref{eqn:pdflambdas}) allows for a novel estimate of 
how redshift space statistics are expected to differ from those 
in real space as $f_{nl}$ varies.  
This is because the overdensity in redshift space is 
\begin{equation}
1 + \delta_s \approx 
 1 + {\delta}_s^{(1)}+ {\delta}_s^{(2)} +{\delta}_s^{(3)}
    + \dots,
\end{equation}
where 
\begin{align}
 \delta_s^{(1)} = &  \delta_r^{(1)} + \Delta_z^{(1)} \nonumber \\
 \delta_s^{(2)} = & \delta_r^{(2)} + \Delta_z^{(2)} 
                   + \delta_r^{(1)}\,\Delta_z^{(1)} \nonumber \\
 \delta_s^{(3)} = & \delta_r^{(3)} + \Delta_z^{(3)}  
                   + \delta_r^{(2)}\,\Delta_z^{(1)} 
                   + \delta_r^{(1)}\,\Delta_z^{(2)}, 
\end{align}
with 
\begin{align}
{\delta_r}^{(1)}  = & \sum_{j=1}^{3} \lambda_j  \nonumber \\
{\delta_r}^{(2)}  = & \frac{\nu_2}{2}\,\delta_l^2 
                            + \frac{\delta_l^2}{3} 
                            - \sum_{j \neq k} \lambda_j\lambda_k \nonumber \\
{\delta_r}^{(3)}  = & \frac{\nu_3}{6}\,\delta_l^3
   + \frac{17}{27}\,\delta_l^3
   - 2\delta_l\,\sum_{j\ne k} \lambda_j\lambda_k 
   - 5\lambda_1\lambda_2\lambda_3,
\end{align}
and
\begin{align}
 \Delta_z^{(1)} = & f_1 \sum_{k=1}^3 \lambda_k\, e_k^2 \nonumber \\
 \Delta_z^{(2)} = & f_1 \sum_{k=1}^3
    \left[\frac{\nu_2}{2}\,\frac{f_2}{f_1}-\frac{4}{3}\right]
      \frac{\delta_l^2}{3}\,e_k^2 
    + f_1^2\,\sum_{k=1}^3 \lambda_k^2\,e_k^2 \nonumber \\
  &  + f_1^2 \sum_{j,k=1}^3 \lambda_j\lambda_k\, e_j^2e_k^2 \nonumber \\
 \Delta_z^{(3)} = & f_1 \sum_{k=1}^3 
  \frac{\delta_l^2}{3}\left[(\frac{\nu_2}{2}\,\frac{f_2}{f_1} 
                             - \frac{4}{3})\,\lambda_k \right.\nonumber \\
&\left.  + (\frac{\nu_3}{6}\,\frac{f_3}{f_1} - \frac{2\nu_2}{3} - \frac{5\nu_2}{6}\,\frac{f_2}{f_1} + 2)\,\delta_l\right]e_k^2 \nonumber \\
& + 2f_1^2\sum_{j,k=1}^3\,\lambda_j\,\left[(\frac{\nu_2}{2}\,\frac{f_2}{f_1}-\frac{4}{3})\,\frac{\delta_l^2}{3} + \lambda_k^2\right]\,e_j^2e_k^2 \nonumber \\
& + f_1^3 \sum_{i,j,k=1}^3 \lambda_i\lambda_j\lambda_k\,e_i^2 e_j^2 e_k^2,
\end{align}
\citep{lamshethred}.
Here, $f_1 = d\ln D_1/da \approx \Omega^{0.55}$ where 
$D_1$ is the linear growth factor,  
$f_2 = d\ln D_2/da$ where $D_2/D_1^2\approx -(3/7)\Omega^{-1/143}$, and 
$\nu_2\approx 1.62$ and $\nu_3\approx 3.93$ are related to the 
spherical evolution model.

Note that setting $f_{nl}\ne 0$ simply changes the values of the 
averages over the $\lambda$s.  Hence, to lowest order, 
\begin{eqnarray}
 \langle \delta_s^2\rangle &\approx& \langle (\delta_s^{(1)})^2\rangle 
   =  \langle (\delta_r^{(1)})^2 + 2\delta_r^{(1)}\Delta_z^{(1)} 
                                + (\Delta_z^{(1)})^2\rangle \nonumber \\
  &=& \sigma^2 + \frac{2f_1}{3}\,\sigma^2
         + \frac{f_1^2}{15} \Bigl\langle(3\delta^2 - 4I)\Bigr\rangle\nonumber\\
  &=& \left(1 + \frac{2}{3}f_1\right)\,\sigma^2
         + \frac{f_1^2}{15}\, \Bigl\langle\frac{5\delta^2}{3} 
                       + \frac{4}{3}(\delta^2 - 3I)\Bigr\rangle \nonumber \\
  &=& \left(1 + \frac{2}{3}f_1 + \frac{f_1^2}{9} 
              + \frac{4f_1^2}{45}\right)\,\sigma^2 \nonumber \\
  &=& \left(1 + \frac{2}{3}f_1 + \frac{f_1^2}{5}\right)\,\sigma^2;
\end{eqnarray}
this is Kaiser's formula, so the relation between real and redshift space 
variance is unchanged from the Gaussian case.  

Of course, $f_{nl}$ matters for the higher order moments. 
The next higher order of the redshift-space variance is 
\begin{eqnarray}
\langle \delta_s^2 \rangle^{(2)} & = & 2\langle \delta_s^{(1)}
                               \delta_s^{(2)}\rangle \nonumber \\
  & =& 2\langle\delta_r^{(1)}\delta_r^{(2)}+ \delta_r^{(1)}\Delta_z^{(2)}
  + {\delta_r^{(1)}}^2 \Delta_z^{(1)} + \Delta_z^{(1)}\delta_r^{(2)}
                                                           \nonumber \\ 
   & & \quad +  \Delta_z^{(1)}\Delta_z^{(2)}  
    + \delta_r^{(1)}{\Delta_z^{(1)}}^2 \rangle \nonumber \\
 & = & 2\frac{\sigma S_3}{6}\sigma^3 \left[3\nu_2  + 
       (\nu_2 +\frac{2}{3})f_1 %
      -\frac{44}{45}f_1^2  + \frac{4}{9}f_1^3  \right. \nonumber \\
  & & \qquad\qquad + \left. \frac{\nu_2}{3}f_1f_2 + \nu_2f_2 \right].
\label{eqn:2ndsig}
\end{eqnarray}

The origin of these terms can be understood as follows.  
When $f_{nl}=0$, then one can think of the three terms in Kaisers 
expression as being due to the density-density, density-velocity and 
velocity-velocity power spectra.  Now, velocities are related to 
first derivatives of the potential, whereas densities are related to 
second derivatives.  So one expects the lowest order corrections 
to the Gaussian result to scale as $f_{nl}$.  
Terms in the first order correction (second equality in equation~\ref{eqn:2ndsig}) 
can be interpreted as $B_{ddd}$, $B_{dvv}$, $B_{ddv}$, $B_{ddv}$, $B_{vvv}$, and 
$B_{dvv}$ respectively, where $B$ denotes bispectra, $d$ and $v$ are density and 
velocity. 
Notice that this first order correction in the redshift variance is of lower order compared 
to the case where $f_{nl}=0$ (and hence $\sigma S_3 = 0$). This is generic for models 
with non-vanishing initial skewness \citep{ptreview}. 

This approach can be extended to estimate the real-redshift large scale relation in 
higher order statistics, for example the bispectrum of galaxies to constrain 
$f_{nl}$ (see for example, \citet{ssz04} or more recently \citet{jk09}).
However  complications arise when one includes  
scale dependent halo/galaxy bias \citep{dalaletal08,slosar08,fnlvincent,dspeak09} and 
the validity of the peak-background split approach in computing halo 
bias \citep{mssBias09}. 
These are beyond the scope of this paper and will be explored in future studies.

\label{lastpage}
\end{document}